2411-7684

# An Optimized Framework to Adopt Computer Laboratory Administrations for Operating System and Application Installations

**Miran Hama Rahim Saeed**

Information Technology Dept.
Computer Science Institute
Sulaimani Polytechnic University
University of Human Development
Sulaimani, Iraq
miran.saeed@spu.edu.iq

**Bryar A. Hassan**

Kurdistan Institution for Strategic Studies and
Scientific Research
Sulaimani Polytechnic University
Sulaimani, Iraq
bryar.hassan@kissr.edu.krd

**Shko M. Qader**

Slemani Governorate
Sulaimani, Iraq
shko@slemani.gov.krd

**Abstract:** *Nowadays, in most of the fields, task automation is area of interest and research due to that manual execution of a task is error prone, time consuming, involving more human resources and focus concerning. In the area of Computer laboratory administration, the old fashioned administration cannot run with today's growth, where the Operating System (OS) and required applications are installed on all the machines one by one. Therefore, a framework for automating Lab administration in regards of Operating Systems and Application installations will be proposed in this research. Affordability, simplicity, usability are taken into major consideration. All the parts of the framework are implemented and illustrated in detail which promotes a great enhancement in the area of Computer Lab Administration.*

## 1. INTRODUCTION

The demand on using computer Lab in Kurdistan region/Iraq universities has increased significantly in recent decades. Most of the collages and institutes have put some Information Technology (IT) related modules to their teaching programs, due to its necessity in university and work life. In addition, students of different departments and stages require varieties of Apps on the Computers, especially in the IT related departments. With the growth of demands, using old fashioned administration of laboratories in terms of OS and Apps installations, where everything is done manually, faces the following difficulties.

- Administrators have to install Operating Systems and several different software programs on the machines one by one, which require plenty of time and also need focuses in order not forget any machine with uninstalled a software.
- When a machine face a software related problem that could not be recovered, a clean installation of OS and Apps take a lot of time.
- Involvement of a mass using of removable storage, which leads movement of Computer Viruses and harmful software between the machines.

However, building a framework to guide the Lab administrators to automate most of the task would be desirable, this is the main aim of this paper. Despite the fact that, there are plenty of software tools and services of different vendors provide automation of some different parts of administration, bringing them up together as a single package with following capability, would completely automate the work of administrator in regard of Oss and Apps deployment.

- All software installation will be done one time on a single machine which reduces spending time and focus significantly for the administrators.
- Barely using removable storage and CD/DVDs, which causes harmful software transfer between the machines.
- The framework will not have any requirement of new hardware, as much as the machines are connected together via a cabled network. On the other hand, with the use of stable, reasonable and easy to use software.
- Reducing routine and automating most of the task.
- Beginner to moderate laboratory administrators will be able to work on the framework.

## 2. METHODOLOGY

The method by which the formwork and its steps can be used will be revealed in this section. Determining the required Apps would be the first step, then choosing a suitable Operating System to give the best performance of the machine and compatible with all the selected Apps will be the next step. After that, a customized OS, which contain all the required Apps, need to be produced as explained in detail in section (3.1), the focus is on Microsoft Windows as it is the most popular Desktop OS according to a statistic published on NETMARKETSHARE™ [1] on June, 2017, more than 90% of Desktop OS users accessed internet using different version of windows. In addition, for related and non-related IT users and departments, Windows is easier to use and mange compared to Linux and Mac. On the other hand, installing Applications on Linux involves a bunch of commands which is not a simple task for administrators. Then, preparing an environment to install the customized Windows on the desired machines via cabled network which has been described deeply in section (3.2). Finally, making a few clicks let the administrators install the customized Windows on any machines connected to the network. Figure (1) illustrates the proposed framework.

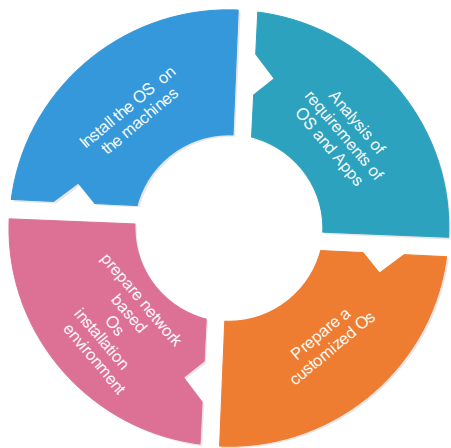

**Figure 1** Proposed framework of Operating Systems and Applications deployment for computer laboratory

## 3. Implementation

In this section, the easiest and most stable methods of producing a customized Windows will be explained in depth firstly. Then, the steps of preparing an environment for remote OS installation will be revealed. Having in mind that, the work is to make administrator's job much easier and simpler, the research has been conducted to find out the most efficient and smooth ways of doing what has been planned for.

### 3.1. Preparing Customized Windows

In this part, all the required steps to create a customized Windows, which include all desired Apps and necessary derivers, will be explained in detail. Firstly, a clean or a fresh (unchanged) Windows 10 need to be installed on a desired machine, which is the newest Desktop Operating System product of Microsoft, and requires the following hardware specifications [2]

Processors: 1 gigahertz (GHz) or faster
RAM: 1 gigabyte (GB) for 32-bit or 2 GB for 64-bit
Hard Disk space: 16 GB for 32-bit OS 20 GB for 64-bit OS.

After the completion of the Windows installation, device drivers and necessary Apps need to be installed. The system image of the partition, where the OS and Apps were installed in it, then has to be backed up on a desired drive which can be done through *Control panel/Backup and Restore*. By completing system back up, a Virtual Hard Disk (VHD) file will be generated, which encapsulates all the contents of the drive and can be mounted as a New Technology File System (NTFS) partition [3]. At this stage, we need to convert the VHD file into Windows Imaging (WIM) file. The WIM file was introduced in Windows Vista and continued in Windows 7, 8 and 10. It is a compressed package that contain related files such as Program files, Windows files and User files [4]. The WIM file can be seen in any extracted Windows image with the name *install.wim* on the *Sources* directory. Once the conversion is made, the name has to be changed to *install* and replaced with an original *install.wim* file of a fresh Windows, hence a customized Windows include all required apps will be achieved.

There are several ways to convert VHD file into WIM file. But firstly, the VHD file has to be mounted as a NTFS partition which can be done through *Computer Management/Disk Management* as explained in Figure (2). Secondly, the WIM file can be generated from the new partition using the following methods

- Command line interface (CLI): by using Windows PowerShell, which is by default available on Windows 8 and 10. It is designed for system administrators to perform complex tasks through the use of task-based command-line shell and scripting language [5]. the following cmdlet syntax let the administrator to generate the WIM file form any partition using Windows PowerShell [6]

*New-WindowsImage -ImagePath <String> CapturePath <String> -Name <String>*

*ImagePath*: the path where the WIM will be saved on
*CapturePath*: the path of the VHD partition, which want to be captured

- Graphical user interface (GUI): by using portable software such as GImageX, which is available on [7], WIM file can be captured easily. At the *Capture* menu of the software, the VHD partition will be browsed as *Source*. The *Destination* is where the WIM file will be saved
on. By clicking on the *Create* button, the software will start converting the VHD file to WIM file.

It is worth mentioning that, the time taken to convert VHD file to WIM image depends on the VHD file size, it could be a couple of minutes to several minutes. When the conversion will be completed successfully, as mentioned before, the name of the WIM image must be change to *install*, then it need to be replaced with *install.wim* file of an extracted fresh Windows image, as a result a Windows with all required applications and drivers will be gotten. The aforementioned process illustrated deeply in Figure (2).

Furthermore, there are other methods let you create a customized *install.wim* image from a fresh one, but each of them has its own draw back and very much depends on the administrator's choice. For example, by using NTlite software tools [8], user easily can load *install.wim* of a fresh Windows, then select all wanted Apps and drivers then generate a customized *install.wim*, hence a customized Windows. Although using NTlite is simple way compared what has been focused in this paper, it only supports single file per App, which means does not support those Apps that has a *setup.exe* file that access a bunch of files outside the *setup.exe* during installation. Furthermore, NTlite runs application installer after Windows setup which requires to be installed by administrators one by one.

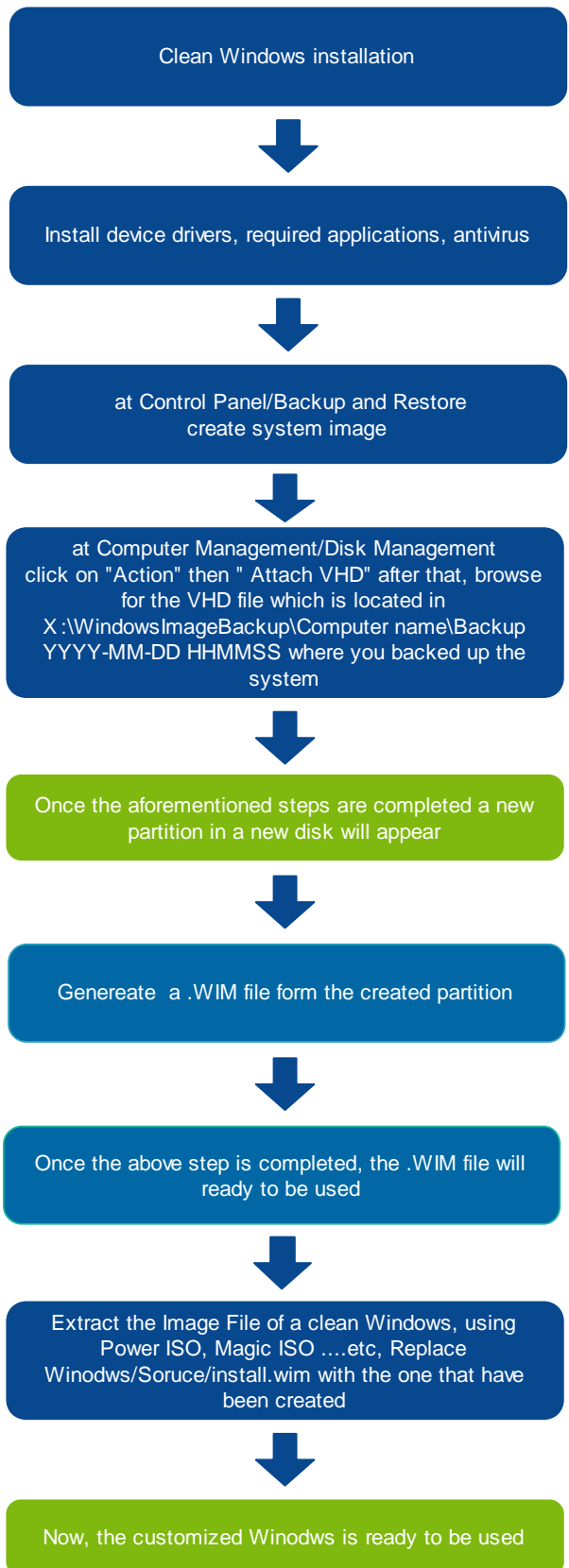

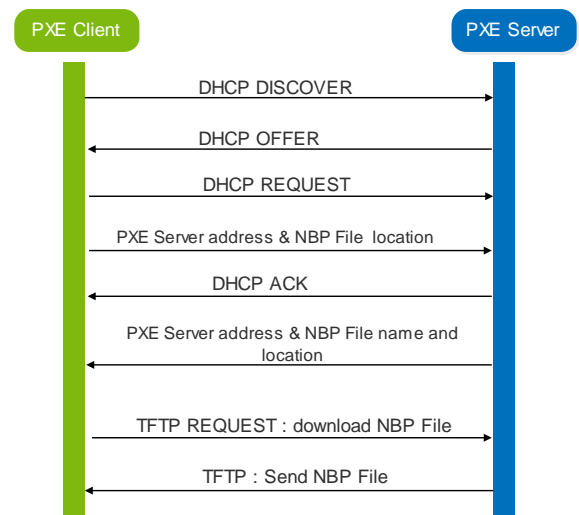

**Figure 3** PXE protocol sequence diagram

## 3.2. Network Based OS Installation

In the previous part, it has been illustrated that, how a customized Windows can be prepared. In this part, the focus will be on the network based installation of the previously customized Windows. The environment of installing OS remotely is a Client/Server model, where the Client needs OS installation remotely and the Server provides all the required files for the client. Accomplishing this task requires Pre-Boot Execution Environment (PXE) Protocol which is designed by intel and implemented in Network Interface Card (NIC) Basic Input/Output System (BIOS) extension or in recent computers (PCs) as part of Unified Extensible Firmware Interface (UEFI) firmware. Basically, PXE enable PCs to boot form server and install OS using Dynamic Host Configuration protocol (DHCP), proxyDHCP, Trivial File Transfer Protocol (TFTP) [9][10]. Figure (3) show the sequence diagram of how PXE protocol works.

After the PXE client receive IP address packet offer, it also broadcast to request for the PXE server IP address and the Network Boot program or Network Bootstrap Program (NBP) file location. the Server then send PXE IP address and the location and name of the NBP file. When the NBP and its configuration downloaded by the client using TFTP, it will be loaded into the Random Access Memory (RAM) and executed, then it starts the boot loader and initialize Windows deployment [10]. However, in multi boot environment, when the NBP executed, a menu will be opened on the client side which shows the available OSs to be booted.

There are plenty ways of doing what have been explained on the previous paragraphs. However, the easiest for LAB administrators and less costly are highly desirable and will be targeted in this paper.

Windows deployment Service (WDS) is one of the Windows Server Role that let administrators to deploy the OS over network, but it has the following disadvantages in our case.

- Cost: licensing of Windows Sever is indeed costly. For instance, the Essential Edition of Windows

**Figure 2** Processes of creating a Customized Windows

Server 2016 costs $501, Which is targeted to small organizations for up to 25 users and 50 devices [11]. In addition, dedicated server also requires a powerful hardware which needs to be taken into consideration.

- Installation and configuration: operating WDS on a server involves lots of work and required a highly skilled administrator. In addition, it has several prerequisites [12]. For those reasons, WDS will not be focused on in this work.

After a comprehensive research, SERVA was found as a most preferable software tool for this purpose. It is an all in one portable PXE server which provides all the required protocols. Serva has been chosen for the following Reasons.

- Portable: Serva does not need installation, it is light and nearly three Megabyte EXE software.
- Reasonable cost: the professional edition v3, which is the last version and compatible with windows 10, costs only $44.99 for only 32bit or 64bit OS architecture. In addition, it provides community edition for free, but with limited capability, which is suitable for home user.
- In dependency: unlike WDS, it does not have any prerequisites
- Ease to use: the interfaces are simple and vey understandable for beginner to moderate administrators [13].

Installation OSs using Server involves few steps of configurations which have been illustrated in figure (4). After installing Serva form [13] on the server side, a directory need to be created for the TFTP server root directory. The Serva then need to be executed, which does not need installation. The following configurations steps need to be done.

- TFTP Server configuration: on the TFTP tab, the machine where the Serva runs on accounted as a TFTP server by Checking TFTP Server check box, then bind the TFTP server IP address with the server IP address, finally browse for the created directory to be as the TFTP server root directory (see appendix A)
- DHCP Server: if there is no DHCP server on the network, on the DHCP tab check the DHCP server box, then check the box Boot Information Negotiation Layer (BINL), which manages a number of PXE client request/response messages to achieve remote OS installation and also checks client credentials. then bind the DHCP server IP address with the server IP address. By adding the DHCP first IP address to be given, subnet mask and the pool size, DHCP configuration will be completed (see appendix B).
- proxy DHCP: in an environment where there is a primary DHCP server, the PXE server does not need to provide IP address for the PXE clients. However, it needs to provide it with the NBP file location and PXE server IP address which is done by proxy DHCP mechanisms. At this case, the check box will be on proxy DHCP and BINL [13] (see appendix C).

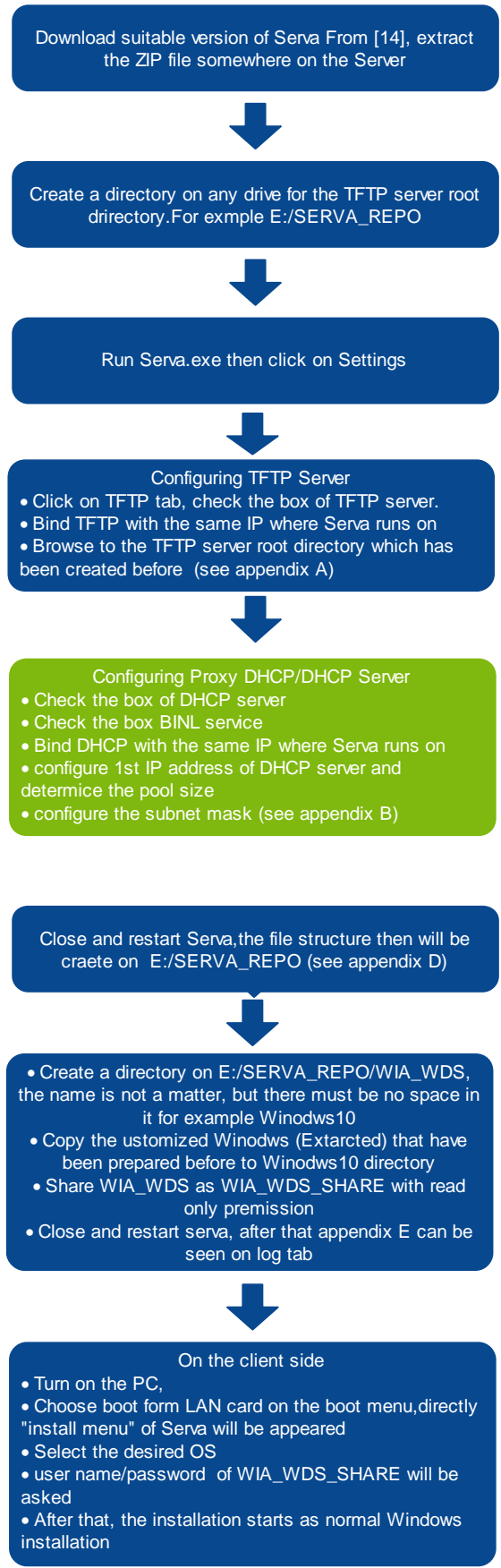

**Figure 4** Processes of preparing Network based OS installation using SERVA

- After the configuration is completed, the Serva need to be restarted, the file structure will be created autmatically on TFTP root directroy (see appendix D). on WIA_WDS directroy create a folder then copy the extracted customized Windows, which has been prepared before. Finally Share WIA_WDS as WIA_WDS_SHARE with read only permission. By closing and restarting Serva, appendix E can be seen, if there are no errors on the configurations

- On the client side a few steps need to be done as explained on the last block of Figure (4).

## 4. DISCSSUION

It is worth mentioning that, the framework is multi boot environment, which means in WIA_WDS directory, administrator can put more than one customized Windows for different purposes and able to install a desired one on the machines. For example, having two various computer type in a Lab requires two distinct customized Windows due to their differences in drivers, or may need to install 32 bit Windows architecture on some machines meanwhile 64 bit architecture on some others based on requirement. This extremely helps administrator to be able to analyze all requirements firstly, then prepare all the customized Operating Systems, so that the work load will be a one-time task. In addition, the Serva does support remote installation for Linux OSs which is out of focus in this paper.

This type of installation may take more time compared to fresh Windows deployments, due to the fact that, the size of customized WIM file is based on the installed Apps in it and clearly bigger than a WIM file of a fresh Windows.

The software (SERVA) is not resource demanding, it can be run on any machine connected to the network.

## 5. CONCLUSION

In this paper, a framework of Operating Systems and Applications installation for computer laboratory has been proposed. The implementation of the framework has been explained thoroughly with usability and affordability in mind. The work promotes automation of Lab administrator's job and it is less error prone compared to the old fashioned ways. The effect clearly can be seen, where there is plenty of machines in a Lab. In addition, the framework does not need a high experienced administrator to follow and also does not require any new hardware but needs a reasonable software. Moreover, implementing the framework eliminate the use of removable storage and its side effects.

Guide for Windows Server 2012", *Technet.microsoft.com*, 2015. [Online]. Available: https://technet.microsoft.com/en-us/library/jj648426(v=ws.11).aspx. [Accessed: 13- May-2017].

# 7. APPENDICES

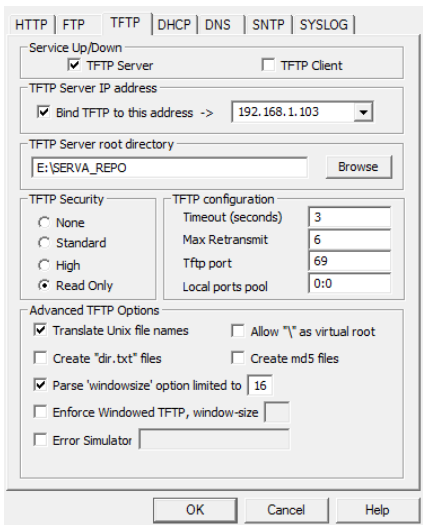

**Appendix A** TFTP server configuration

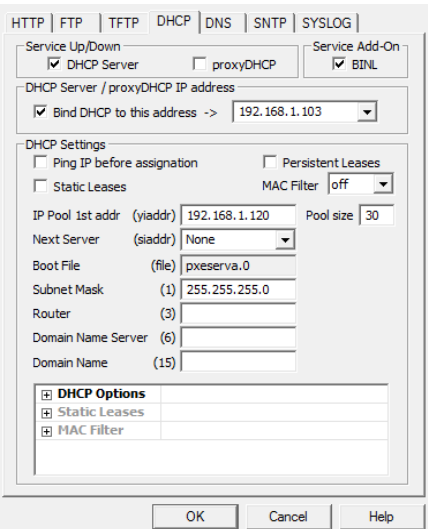

**Appendix B** DHCP server configuration

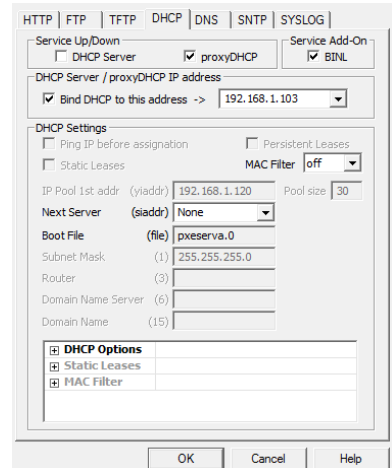

**Appendix C** proxyDHCP server configuration

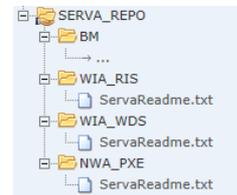

**Appendix D** SERVA file structure

**Appendix E** Final configuration log